\documentclass[structabstract]{aa}  
\usepackage{graphicx}
\usepackage{natbib}
\bibpunct[]{(}{)}{;}{a}{}{,}
\usepackage{txfonts}
\begin{document}
\title{Brown dwarfs and very low mass stars in the Praesepe open
  cluster: a dynamically unevolved mass function?\thanks{Based in part on
    observations carried out at ESO/La Silla, Chile under proposal
    number 078.A-9055(A).}}  \author{S. Boudreault\inst{1}, C. A. L.
  Bailer-Jones\inst{1}, B. Goldman\inst{1}, T. Henning\inst{1} and J.
  A. Caballero\inst{2}}

\institute{Max-Planck-Institut f\"ur Astronomie,
  K\"onigstuhl 17, D-69117 Heidelberg, Germany\\
  \email{boudrea, calj, goldman, henning@mpia.de} \and
  Departamento de Astrof\'isica, Facultad de F\'isica, Universidad
  Complutense de Madrid, E-28040 Madrid, Spain}

\date{Received 30 July, 2009; accepted 23 October, 2009}

 
\abstract
{Determination of the mass functions of open clusters of different ages
  allows us to infer the efficiency with which brown dwarfs are
  evaporated from clusters to populate the field.}
{In this paper we present the results of a photometric survey to
  identify low mass and brown dwarf members of the old open cluster
  Praesepe (age 590$^{+150}_{-120}$\,Myr, distance
  190$^{+6.0}_{-5.8}$\,pc) from which we estimate its mass function and compare this with that of other clusters.}
{We performed an optical ($I_{\rm c}$-band) and near-infrared
  ($J$ and $K_{\rm s}$-band) photometric survey of Praesepe covering
  3.1\,deg$^2$. With 5$\sigma$ detection limits of
  $I_{\rm c}=23.4$ and $J=20.0$, our survey is predicted to be sensitive to objects
  with masses from 0.6 to 0.05\,M$_\odot$.}
{We photometrically identify 123 cluster member candidates
  based on dust-free atmospheric models and 27 candidates
  based on dusty atmospheric models.  The mass function
  rises from 0.6\,M$_\odot$ down to 0.1\,M$_\odot$ (a power law fit of
  the mass function gives $\alpha$=1.8$\pm$0.1;
  \,$\xi$(M)\,$\propto$\,M$^{-\alpha}$\,), and then turns over at
  $\sim$0.1\,M$_\odot$. This rise agrees with the
  mass function inferred by previous studies, including a survey based
  on proper motion and photometry. In contrast, the mass function differs significantly from that measured for the Hyades, an open cluster with a similar age
  ($\tau$\,$\sim$\,600\,Myr).
  Possible reasons are that the clusters did
  not have the same initial mass function, or that dynamical
  evolution (e.g.\ evaporation of low mass members) has proceeded differently in the two clusters. Although different binary fractions could
  cause the observed (i.e.\ system) mass functions to differ,
there is no evidence for differing binary fractions from 
  measurements published in the literature.  Of our
  cluster candidates, six have masses predicted to be equal to or
  below the stellar/substellar boundary at 0.072\,M$_\odot$.}
{}

\titlerunning{Brown dwarfs and very low mass stars in the Praesepe
  cluster}

\authorrunning{Boudreault, S., et al.}

\maketitle

\section{Introduction}

Several publications in the past decade have been concerned with the
mass function (MF) of low mass stellar and substellar populations in
open clusters, including $\sigma$~Orionis (\citealt{bejar2002},
\citealt{caballero2007}), the Orion Nebula Cluster
(\citealt{hillenbrand2000}, \citealt{slesnick2004}), IC~2391
(\citealt{barrado2004}, \citealt{boudreault2009}), the Pleiades
(\citealt{moraux2003}, \citealt{lodieu2007}), and the Hyades
(\citealt{reid99}, \citealt{bouvier2008}), to name just a few.
Studies of relatively old open clusters
(age\,$\gtrsim$\,100\,Myr) are important for the following two reasons
in particular. First, they permit a study of the intrinsic evolution
of brown dwarfs (BDs), e.g.\ their luminosity and effective
temperature, which constrains structural and atmospheric models.
Second, together with younger clusters we can investigate how BD
populations as a whole evolve and thus probe the efficiency with which BDs
evaporate from clusters to populate the Galactic field.  Numerical
simulations of cluster evolution have demonstrated that the MFs can
evolve through dynamical interaction (\citealt{marcos2000};
\citealt{adams2002b}).  These interactions result in a decrease of the
open cluster BD (and low-mass star) population.  This has been
observed by \cite{bouvier2008} from a comparison of the Pleiades
(120\,Myr) and Hyades (625\,Myr) mass functions.

Many earlier studies of the substellar MF have focused on young open
clusters with ages less than $\sim$\,100\,Myr, and in many cases much
younger ($< 10$\,Myr). This is partly because BDs are bright when they
are young (lacking a significant nuclear energy source, they cool as
they age), thus easing detection of the least massive objects.
However, youth presents difficulties. First, intra-cluster extinction
plagues the determination of the intrinsic luminosity function from
the measured photometry. Second, at these ages the BD models have
large(r) uncertainties (\citealt{baraffe2002}). Estimates of the substellar MF
in very young clusters (age $\lesssim$1\,Myr) might be
unreliable due to these modelling uncertainties
(\citealt{chabrier2005}). BDs in older clusters suffer less from
these problems, but have the disadvantage that much deeper surveys are
required to detect them.

The old open cluster Praesepe is an interesting target considering its
age and distance. It is located at a distance of
190$^{+6.0}_{-5.8}$\,pc (based on parallax measurements from the new
Hipparcos data reduction, \citealt{Leeuwen2009}) and has an age of
590$^{+150}_{-120}$\,Myr (by isochrone fitting in the
Hertzsprung-Russell diagram; \citealt{fossati2008}). The extinction
towards this cluster is low, $E(B-V)$\,=\,0.027$\pm$0.004\,mag
(\citealt{taylor2006}), while determinations of the metallicity of
Praesepe yield some discrepancies: [Fe/H]\,=\,0.038$\pm$0.039,
\cite{friel1992}; +0.13$\pm$0.10, \cite{boesgaard1988}; 0.11$\pm$0.03
from spectroscopy and 0.20$\pm$0.04 from photometry, \cite{an2007};
+0.27$\pm$0.10, \cite{pace2008}.  \cite{hambly1995} presented a
$\sim$19\,deg$^2$ survey of the Praesepe cluster down to masses of
$\sim$0.1\,M$_\odot$ and observed a rise of the MF at the lowest
masses. They concluded that this implied a large
population of BDs. A shallow survey complete to $I$\,=\,21.2\,mag,
$R$\,=\,22.2\,mag over 800\,arcmin$^2$ uncovered one spectrally
confirmed very low-mass star or BD (spectral type of M8.5V) with a
model-dependent mass of 0.063--0.084\,M$_\odot$ (\citealt{Ma_98}). A
survey over the central 1\,deg$^2$ with 10$\sigma$ limits of
$R$\,=\,21.5, $I$\,=\,20.0 and $Z$\,=\,21.5\,mag revealed 19~BD
candidates and the first MF determination of Praesepe down to the
substellar limit, but without spectral confirmation
(\citealt{pinfield97}). Subsequent infrared photometry of the sample reduced this
number to nine candidates (\citealt{hodgkin99}).  \cite{adams2002a}
presented a 100\,deg$^2$ study of Praesepe using 2MASS (Two-Micron All
Sky Survey) data and Palomar survey photographic plates, from which they
derived proper motions. They determined the radial profile of this
cluster but their MF does not reach the substellar regime.  A more
recent proper motion survey of Praesepe covers a much larger area
(300\,deg$^2$; \citealt{kraus2007}), but does not reach the BD regime
either (the limit is $\sim$0.12\,M$_\odot$). Finally, the most recent
substellar MF determination of Praesepe was published by
\cite{gonzalez-garcia2006} and extends to a 5$\sigma$ detection limit
of $i=24.5$\,mag corresponding to 0.050--0.055\,M$_\odot$.  They
identified one new substellar candidate, but their survey covers only
1177 arcmin$^2$.

In this paper, we present the results of a program to study, in
detail, the MF of Praesepe down to the substellar
regime. Our photometric survey is, as with \cite{gonzalez-garcia2006},
the deepest so far in optical and near-infrared (NIR) bands, with
5$\sigma$ detection limits of $I_{\rm c}=23.4$ and $J=20.0$
(corresponding to a mass limit of about 0.05\,M$_\odot$), but covers
more than nine times the area.  Our paper is structured as follows. We
first present the data set, reduction procedure and calibration in
section \ref{obs-data-calib}.  We then discuss our candidate selection
procedure in section \ref{selection} and the survey results in section
\ref{results-survey} before discussing the derived MF in section
\ref{mf-variation}. We conclude in section \ref{conclusions}.

\section{\label{obs-data-calib} Observations, data reduction,
  calibration, and estimation of masses and effective temperatures}

\subsection{\label{obs} Observations}

\begin{table}
  \caption{Description of observations with the O2k infrared camera.}
  \label{tab:observations_o2k}
  \centering
  \begin{tabular}{ccccccc}
    \hline\hline
    Field & $\alpha$ & $\delta$ & t$_{\rm exp}$($J$) & t$_{\rm exp}$($K_{\rm s}$) & $J$(5$\sigma$) & $K_{\rm s}$(5$\sigma$) \\ 
    & ($^h$\,$^m$\,$^s$) & ($^\circ$\,$'$\,$''$) & [min] & [min] & [mag] & [mag] \\
    \hline
    001&08 40 04&+19 40 00& 60& 40&19.9 &18.6 \\
    A01&08 41 04&+19 54 00& 60&130&20.6 &19.4 \\
    A02&08 40 04&+19 54 00& 40& 40&19.5 &18.5 \\
    A03&08 39 04&+19 54 00& 40& 40&19.7 &18.7 \\
    A04&08 39 04&+19 40 00& 20& 40&19.0 &18.6 \\
    A05&08 39 04&+19 26 00& 20& 40&20.3 &18.8 \\
    A06&08 40 04&+19 26 00& 20& 40&19.9 &18.6 \\
    A07&08 41 04&+19 26 00& 40& 60&20.3 &19.0 \\
    A08&08 41 04&+19 40 00& 20& 40&18.0 &17.6 \\
    B01&08 42 04&+20 08 00& 40& 40&19.7 &18.4 \\
    B02&08 41 04&+20 08 00& 40& 40&18.8 &18.6 \\
    B03&08 40 04&+20 08 00& 20& 40&20.0 &18.5 \\
    B04&08 39 04&+20 08 00& 20& 40&19.7 &18.6 \\
    B05&08 38 04&+20 08 00& 20& 40&20.3 &18.7 \\
    B06&08 38 04&+19 54 00& 20& 40&20.0 &18.5 \\
    B07&08 38 04&+19 40 00& 20& 40&19.5 &18.5 \\
    B08&08 38 04&+19 26 00& 20& 40&17.4 &17.8 \\
    B09&08 38 04&+19 12 00& 20& 40&18.0 &18.8 \\
    B10&08 39 04&+19 12 00& 20& 40&20.1 &18.8 \\
    B13&08 42 04&+19 12 00& 20& 40&20.1 &18.8 \\
    B14&08 42 04&+19 26 00& 20& 40&20.1 &18.7 \\
    B15&08 42 04&+19 40 00& 20& 40&19.3 &18.3 \\
    B16&08 42 04&+19 54 00& 20& 40&18.4 &18.1 \\
    C01&08 43 04&+20 22 00& 20& 40&20.4 &18.8 \\
    C02&08 42 04&+20 22 00& 20& 40&20.1 &18.7 \\
    C04&08 40 04&+20 22 00& 20& 40&20.0 &18.8 \\
    C05&08 39 04&+20 22 00& 20& 40&19.1 &18.2 \\
    C06&08 38 04&+20 22 00& 20& 40&20.1 &18.7 \\
    C07&08 37 04&+20 22 00& 20& 40&19.8 &18.2 \\
    C08&08 37 04&+20 08 00& 20& 40&19.9 &18.7 \\
    C09&08 37 04&+19 54 00& 20& 40&20.1 &18.0 \\
    C10&08 37 04&+19 40 00& 20& 40&20.3 &19.3 \\
    C11&08 37 04&+19 26 00& 20& 40&19.6 &18.6 \\
    C12&08 37 04&+19 12 00& 20& 40&20.3 &18.9 \\
    C13&08 37 04&+18 58 00& 20& 40&20.4 &18.7 \\
    C14&08 38 04&+18 58 00& 20& 40&19.9 &18.4 \\
    C15&08 39 04&+18 58 00& 20& 40&20.4 &18.6 \\
    C16&08 40 04&+18 58 00& 20& 40&19.4 &18.4 \\
    C17&08 41 04&+18 58 00& 20& 40&19.3 &18.3 \\
    C18&08 42 04&+18 58 00& 20& 40&20.6 &18.3 \\
    C19&08 43 04&+18 58 00& 20& 40&20.3 &18.4 \\
    C20&08 43 04&+19 12 00& 20& 40&19.6 &18.6 \\
    C21&08 43 04&+19 26 00& 20& 40&20.3 &18.8 \\
    C22&08 43 04&+19 40 00& 20& 40&20.3 &18.7 \\
    C23&08 43 04&+19 54 00& 20& 40&19.8 &18.7 \\
    C24&08 43 04&+20 08 00& 20& 40&19.1 &18.4 \\
    \hline
  \end{tabular}
\end{table}

\begin{table}
  \caption{Description of observations with WFI optical camera.}
  \label{tab:observations_wfi}
  \centering
  \begin{tabular}{ccccc}
    \hline\hline
    Field & $\alpha$ & $\delta$ & t$_{\rm exp}$($I_{\rm c}$) & $I_{\rm c}$(5$\sigma$) \\ 
    & ($^h$\,$^m$\,$^s$) & ($^\circ$\,$'$\,$''$) & [min] & [mag] \\
    \hline
    1&08 40 04&+19 40 00& 24&22.6 \\
    2&08 42 24&+20 15 00& 30&23.3 \\
    3&08 40 04&+20 15 00& 36&23.2 \\
    4&08 37 44&+20 15 00& 24&23.4 \\
    5&08 37 44&+19 40 00& 36&23.6 \\
    6&08 37 44&+19 05 00& 36&23.1 \\
    7&08 40 04&+19 05 00& 36&23.5 \\
    8&08 42 24&+19 05 00& 36&22.8 \\
    9&08 42 24&+19 40 00& 36&22.8 \\
    \hline
  \end{tabular}
\end{table}

Our survey consists of 47 Omega 2000 (O2k) fields each of size
15.4$\times$15.4 arcmin$^2$ observed in $J$ and $K_{\rm s}$, plus the
same region observed in nine $I_{\rm c}$ Wide Field Imager (WFI)
fields each of size 34$\times$33 arcmin$^2$. This gives a total
coverage of 3.1\,deg$^2$ observed in all three bands, centred on
RA(J2000)=08$^h$40$^m$04$^s$ and DEC(J2000)=+19$^\circ$40$'$00$''$.

\begin{figure}
  \includegraphics[width=8cm]{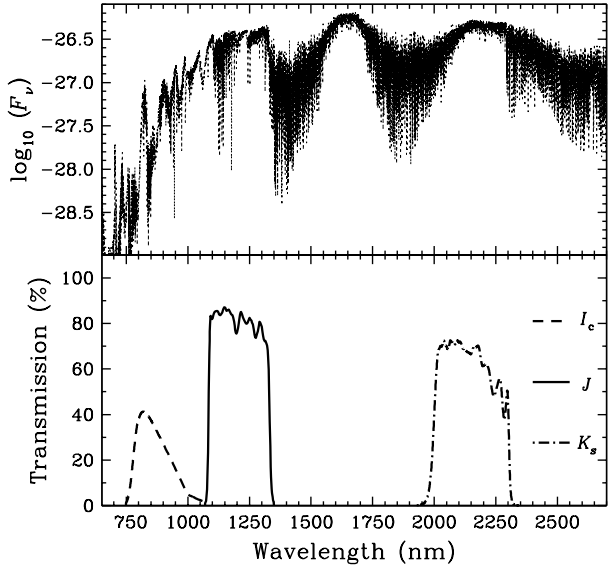}
  \caption{\label{fig:passband} Transmission curve of the filters used
    in our survey compared to the synthetic spectrum of a BD with
    $T_{\rm eff}$ = 2300 K, log \textit{g} = 4.5 and solar metallicity
    (NextGen model). The transmission curves include the quantum
    efficiency of the detectors.}
\end{figure}

The near-infrared (NIR) observations were made on the 3.5m telescope at Calar Alto,
Spain (with observation runs of several nights from February 2005 to
January 2007). O2k (\citealt{calj2000}; \citealt{baumeister03})
comprises a HAWAII-2 detector with 2k$\times$2k pixels over a field
of view of 15.4$\times$15.4 arcmin delivering a pixel scale of 0.45\,arcsec
per pixel.  The optical observations were carried out with the Wide
Field Imager (WFI) on the MPG/ESO 2.2m telescope at La Silla, Chile
(\citealt{baade1999}) during 17--22 March 2007.  The WFI is a mosaic
camera of 4$\times$2 CCDs, each with 2k$\times$4k pixels,
covering a total field of view of 34$\times$33 arcmin$^2$ at
0.238\,arcsec per pixel. All fields were observed in the broad band
filter $I_{\rm c}$. A detailed list of the fields observed with
pointing, filter, exposure time and 5$\sigma$ detection limits is
given in Table \ref{tab:observations_o2k} for the NIR data and in
Table \ref{tab:observations_wfi} for the optical data. The passband
functions for the filters, multiplied with the quantum efficiency of
the detectors, are shown in Figure \ref{fig:passband}.

\subsection{\label{data} Data reduction and astrometry}

The standard data reduction steps (overscan subtraction, trimming and
flat-fielding for the WFI data; dark subtraction and flat-fielding for
O2k data) were performed on a nightly basis, using the $ccdred$
package under IRAF\footnote{The O2k camera suffers from a stray light
  problem. It appears on every image taken with the camera,
  forming a ring pattern centred in the middle of the detector
  (\citealt{nicol09}). The stray light is removed via our global
  illumination correction and sky subtraction.}. For both WFI and O2k
data we used \textit{superflats} (obtained by combining science image
frames for each night) for pixel-to-pixel variation correction and for
correcting the global illumination. For our NIR data, the sky
background was subtracted using the median-combined images for each
filter and each field (on a nightly basis).  For WFI data, we reduced
each of the eight CCDs in the mosaic independently and in the final
step scaled them to a common flux response level.  We made an initial
sky subtraction via a low-order fit to the optical data, and for the
infrared data by subtracting a median combination of all
(unregistered) images of the science frames.  Fringes were visible for
the $I_{\rm c}$-band photometry. They were removed in the way
described by \cite{calj2001}\footnote{A \textit{fringe correction
    frame} was created, which is a median combination of all science
  frames in a same filter with the same exposure time.  This
  \textit{fringe correction frame} was scaled by a factor, determined
  manually for each science frame, and subtracted from the science
  image.}.  Finally, the individual images of a given field were
registered and median combined. We used the IRAF task \textit{daofind}
to automatically detect stellar objects in an image by approximating
the stellar point spread function with a Gaussian. We
visually inspected the images in order to remove from our cluster
candidate list any extended sources (i.e.\ galaxies) that were
mistakenly identified as stars by \textit{daofind} (see section
\ref{selection-3rd}). Sources were extracted and instrumental
magnitudes assigned via aperture photometry with the IRAF task
\textit{wphot}. To this aperture photometry we have applied an
aperture correction following the technique described in \cite{howell1989}.
An astrometric solution was obtained using the IRAF package
\textit{imcoords} and the tasks \textit{ccxymatch}, \textit{ccmap} and
\textit{cctran}. For each WFI field, this solution was computed for
the $I_{\rm c}$-band image (and for each O2k field using the $J$-band
image) using the 2MASS catalogue as a reference.  The root mean square
accuracy of our astrometric solution is 0.15--0.20 arcsec for both WFI
and O2k data.  For WFI data, the astrometry was performed on a
CCD-by-CCD basis.

\subsection{\label{calib} Photometric calibration}

To correct for Earth-atmospheric absorption on the photometry, we
calibrated the infrared data using the $J$ and $K_{\rm s}$-band magnitudes
of 2MASS objects which were observed in our science fields. By
determining a constant offset between the magnitude of 2MASS and our
instrumental magnitude, we obtained the zero point offset. Since this
zero point offset was obtained with objects in the same field of view
in each science frame, and since we found the difference between the
2MASS and O2k passbands to be insignificant, we did not need to
perform an airmass or colour correction when reducing our NIR
photometry. (That is, the determined coefficients were statistically
consistent with zero.)

We followed a similar approach for our $I_{\rm c}$-band photometry,
but using observations in our fields for which $r$ and $i$-band
magnitudes are available in the Sloan Digital Sky Survey (SDSS)
catalogue. We first transformed the $i$-band magnitudes of SDSS to
$I_{\rm c}$-band magnitudes using the transformation equation of
\cite{jordi2006} \begin{equation}
  I_{\rm c}=i-0.381-0.254\times(r-i)
\end{equation}
We then determined the zero point offset between this $I_{\rm c}$
magnitude and our instrumental $I_{\rm c}$ magnitude, again using the
SDSS stars. A further colour correction was not necessary, and as this
calibration is applied on a field-by-field basis (as with the NIR
data), an airmass correction was likewise unnecessary.

\subsection{\label{get-mass} Mass and effective temperature estimates based on photometry}

After we identify candidates (section 3) we will use the multiband
photometry to derive their masses and effective temperatures, $T_{\rm
  eff}$. We use the evolutionary tracks from \cite{baraffe1998} and
atmosphere models from \cite{hauschildt1999a} (assuming a dust-free
atmosphere; the NextGen model) to compute an isochrone for Praesepe
for an age of 590\,Myr, a distance of 190\,pc, a solar metallicity
and assuming zero extinction.  These models and assumptions provide us
with a prediction of $f_{\lambda}$, the spectral energy distribution
received at the Earth (in erg cm$^{-2}$ s$^{-1}$ \AA$^{-1}$) from the
source. We need to convert these spectral energy distributions into magnitudes in the filters
we used.  Denoting as $S_{A}$($\lambda$) the (known) total
transmission function of filter $A$ (including the CCD quantum
efficiency and assuming telescope and instrumental throughput are
flat), then the flux measured in the filter is
\begin{equation}
  f_{A}=\frac{\int_{0}^{\infty}f_{\lambda}S_{A}(\lambda)d\lambda}{\int_{0}^{\infty}S_{A}(\lambda)d\lambda},
  \label{eqw:phot-model}
\end{equation}
\noindent The corresponding magnitude $m_{A}$ in the Johnson
photometric system is given by
\begin{equation} {m_{A}=-2.5\;\mathrm{log}\;f_{A}\;+\;c_{A}},
  \label{eqw:phot-cal}
\end{equation}
\noindent where $c_{A}$ is a constant (zero point) that remains to be
determined in order to put the model-predicted magnitude onto the
Johnson system.  We derived this constant for each of the bands $I_{\rm
  c}$, $J$ and $K_{\rm s}$ in the standard way, namely by requiring
that the spectrum of Vega produce a magnitude $m_{A}$ of 0.03 in all
bands. Using the Vega spectrum from \cite{colina1996} we derive values
of $c_{I_{\rm c}}=-22.6011$, $c_{J}=-23.6865$ and $c_{K_{\rm
    s}}=-25.9076$\,mag.  Applying the two equations above to a whole
set of model spectra produces a theoretical isochrone in
colour--magnitude space.  Note that this procedure only provides us
with the ``true'' magnitudes of the model spectra, not their
instrumental ones. The photometric calibration procedure applied to
the data converts the measured, instrumental magnitudes to the
``true'' magnitude plane where we then compare them with the isochrone.

Assuming that all our photometric candidates belong to Praesepe, we
derive masses and effective temperatures from these isochrones using
our three filter measurements in the following way. We first normalize
the energy distribution of each object to the energy distribution of
the model using the $J$ filter. We then estimate the mass and
effective temperature via a least squares fit of the measured spectral
energy ``distribution'' (actually just two points) to the model
spectral energy distribution from the isochrone. This involves
estimating one parameter from two measurements, because mass and
$T_{\rm eff}$ are not independent.

The above assumption of a dust-free atmosphere is valid for $T_{\rm
  eff}$~$\geqslant$~3000\,K, but objects with
3000\,K~$\geqslant$~$T_{\rm eff}$~$\geqslant$~1800\,K are expected to
have dust in equilibrium with the gas phase (\citealt{allard01}).  We
therefore perform a second selection of candidates (and determination
of mass and $T_{\rm eff}$) based on isochrones predicted in the same
way, but based on evolutionary tracks of \cite{chabrier2000} and the
AMES-dusty model of \citet{allard01}.  This give us a second
\textit{dusty model} list of candidates. A priori some observed stars
could appear in both lists (and in fact two do), but in our later
discussions of the mass function we do not mix stars from the two
lists but rather make separate determinations of the mass function.

There are various sources of error in the estimation of mass
and $T_{\rm eff}$. These are the photon noise, the photometric
calibration, the least squares fitting (imperfect model) and the
uncertainties in the age of and distance to Praesepe. The uncertainties
in the age and distance are the most significant errors
and given rise to uncertainties of
0.060$\pm$0.010\,M$_\odot$ and 1\,990$\pm$260\,K for a
substellar object, 0.072$\pm$0.008\,M$_\odot$ and 2\,293$\pm$201\,K
for an object at the hydrogen burning limit and
1.000$\pm$0.017\,$M_\odot$ and 5\,300$\pm$50\,K for a solar-type
star.

\section{\label{selection} Candidate selection procedure}

The candidate selection procedure for BDs and very low-mass stars is
as follows (and explained in more detail in the remainder of this
section).  Candidates were first selected based on colour-magnitude
diagrams (CMDs) and this further refined using colour-colour
diagrams. In the third and final selection, we used the known distance
to Praesepe to reject objects based on a discrepancy between the
observed magnitude in $J$ and the magnitude in this band computed with
the isochrones and our estimation of $T_{\rm eff}$. To be considered
as a cluster member, an object has to satisfy all three of these
criteria. We make two independent selections: one using dust-free and
one using dusty atmospheric models.

\subsection{\label{selection-1st} First candidate selection step: colour--magnitude diagrams}

\begin{figure}
  \centering
  \includegraphics[width=8cm]{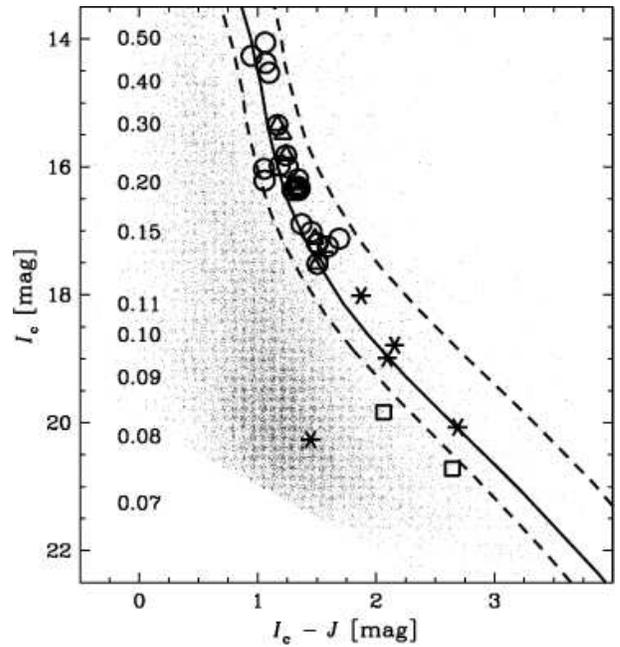}
  \caption{\label{fig:cmd-ij} Colour--magnitude diagram showing an example of 
  the first selection step using the $I_{\rm c}$ and $J$ bands. 
    The solid lines show the isochrone computed from an evolutionary
    model with a dust-free atmosphere (NextGen model) and the dashed lines show
    our selection band around this. The numbers
    indicate the masses (in M$_{\odot}$) of objects on the isochrone for
    various $I_{\rm c}$ magnitudes.  Overplotted are measurements from our survey of 
   candidate cluster members reported in \cite{pinfield97}
    (\textit{stars}), \cite{adams2002a} (\textit{triangles}) [where we
    include objects which have a probability of being a real member
    higher than 10\%], \cite{gonzalez-garcia2006} (\textit{squares})
    and \cite{kraus2007} (\textit{circles}).}
\end{figure}

\begin{figure}
  \centering
  \includegraphics[width=8cm]{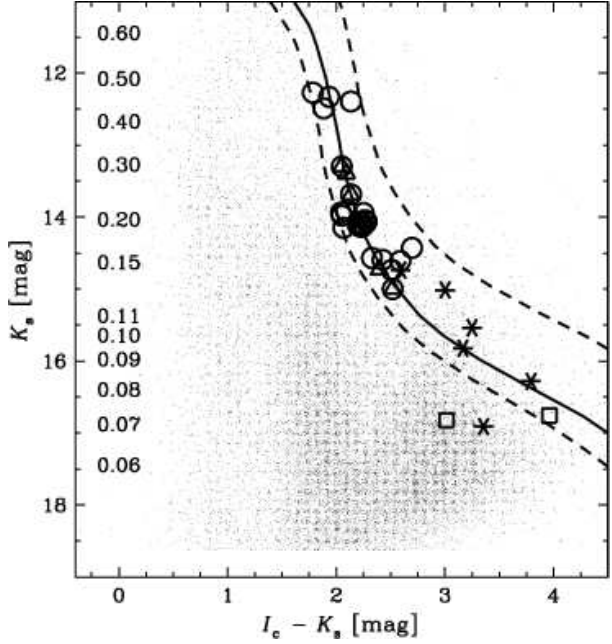}
  \caption{\label{fig:cmd-ik} As Figure \ref{fig:cmd-ij} but
    with the $I_{\rm c}$ and $K_{\rm s}$ bands.}
\end{figure}

Candidates were first selected from our CMDs by retaining only objects
which are no more than 0.14 mag redder or bluer than the isochrone in
all CMDs. This number accommodates errors in the magnitudes,
uncertainties in the model isochrones plus uncertainties in the
cluster age and distance estimates. We additionally include objects
brighter than the isochrones by 0.753\,mag in order to include
unresolved binaries. In Figure~\ref{fig:cmd-ij} and \ref{fig:cmd-ik}
we show two CMDs where candidates were selected based on $I_{\rm c}$
vs. $I_{\rm c}$--$J$ and $K_{\rm s}$ vs. $I_{\rm c}$--$K_{\rm s}$.
These figures also show low-mass cluster member candidates from
previous studies which we detected in our survey
(\citealt{pinfield97}; \citealt{adams2002a};
\citealt{gonzalez-garcia2006}; \citealt{kraus2007}).  In Figure
\ref{fig:cmd-ik}, we can observe three structures in this CMD.  The
two structures at $I_{\rm c}-K_{\rm s}$$\sim$1\,mag and $I_{\rm
  c}-K_{\rm s}$$\sim$2\,mag are predominantly stars (Galactic
disk turn-off, and disk late-type and giant stars respectively) while
the structure at $I_{\rm c}-K_{\rm s}$$\sim$3\,mag is mostly composed
of galaxies.  From a total of 23\,891 objects detected above the
5$\sigma$ detection limit in all filters, 800 are retained as
candidate cluster members (96.7\% are rejected). If we instead use
dusty model isochrones, then out of the 23\,891 objects, 357 are
retained (98.5\% are rejected) for our dusty model list.

\subsection{\label{selection-2nd} Second candidate selection step:
  colour--colour diagram}

\begin{figure}
  \centering
  \includegraphics[width=8cm]{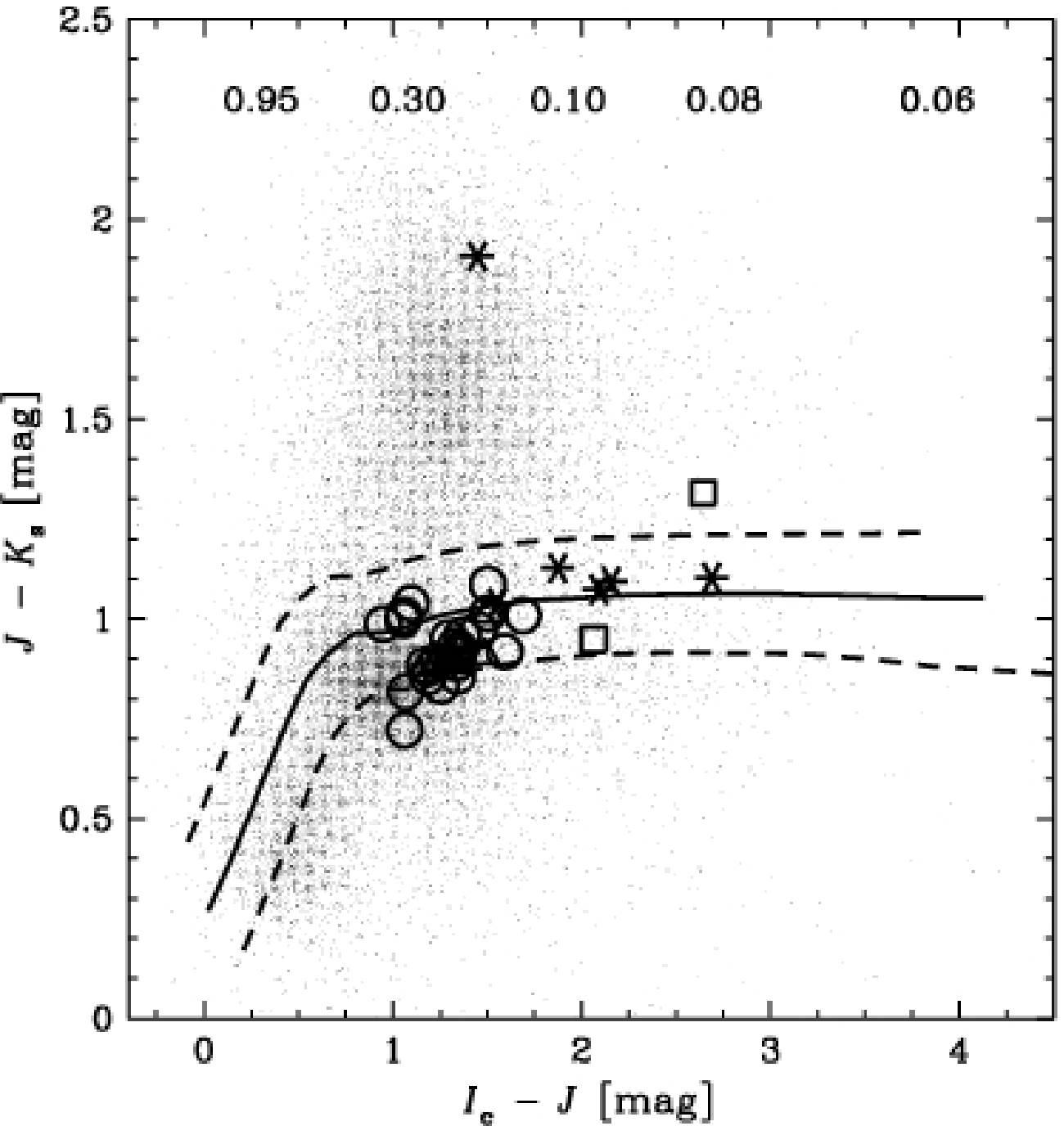}
  \caption{\label{fig:ccd1} Colour-colour diagram used in the second selection step. 
    The solid line is the isochrone computed from an evolutionary
    model with a dust-free atmosphere (NextGen model, the masses in
    M$_{\odot}$ for each $I_{\rm c}-J$ colours are shifted up
   clarity) and the dashed lines show our selection band.
   Overplotted are the cluster candidate members from
    \cite{pinfield97} (\textit{stars}), \cite{adams2002a}
    (\textit{triangles}), \cite{gonzalez-garcia2006}
    (\textit{squares}) and from \cite{kraus2007} (\textit{circles}),
    which we detected in our survey.}
\end{figure}

\begin{figure}
  \centering
  \includegraphics[width=8cm]{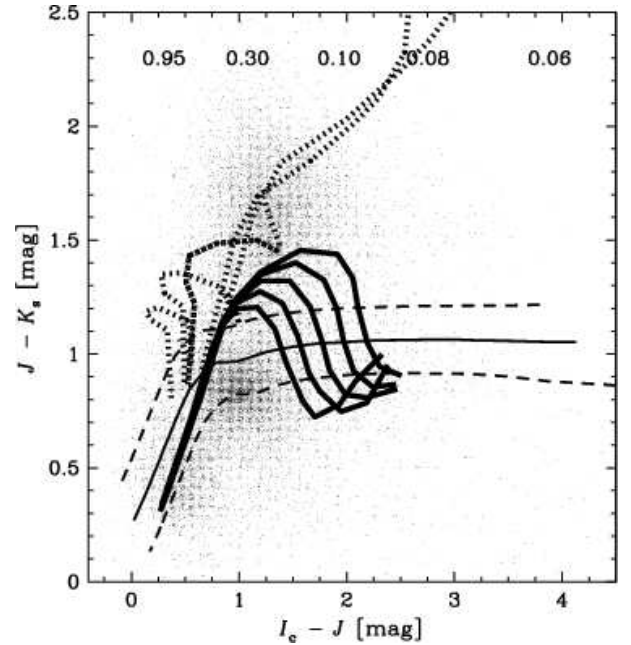}
  \caption{\label{fig:ccd2} As Figure \ref{fig:ccd1}, but now showing
   the theoretical colours of six galaxies as thick dotted lines
  and the theoretical colours of red giants as thick solid lines.
  The six galaxies are two starbursts, one Sab, one Sbc, and two
    ellipticals of 5.5 and 15\,Gyr, with redshifts from $z$=0 to $z$=2
    in steps of 0.25 (evolution not considered).  We assume that all
    red giants have a mass of 5\,M$_{\odot}$, 0.5 $<$ log \textit{g}
    $<$ 2.5 and 2000\,K $<$ $T_{\rm eff}$ $<$ 6000\,K.}
\end{figure}

The second stage of candidate selection involves retaining just those
objects which lie within 0.24 mag of the isochrone in the (single)
colour-colour diagram.  This value reflects the photometric errors
plus uncertainty in the age estimation of Praesepe.  One such
colour-colour diagram with the selection limits is shown in Figure
\ref{fig:ccd1}. The two main sources of contamination beside field M
dwarfs are background red giants and unresolved galaxies (Praesepe is
at a Galactic latitude of $b$=+32.5$^{\circ}$). We show in Figure
\ref{fig:ccd2} the theoretical colours for red giants using the
atmosphere models of \cite{hauschildt1999b} and theoretical colours of
six galaxies from \cite{meisenheimer09}. We see that red giants
could be a source of contamination in the mass range of
0.09--0.2\,M$_{\odot}$ and at $\sim$0.7\,M$_\odot$, while unresolved
galaxies should not be a major source of contamination below
0.6\,M$_{\odot}$. In Figure \ref{fig:ccd2} we see the same three
structures as in Figure \ref{fig:cmd-ik}: from top to bottom galaxies,
disk late-type and giant stars, and Galactic disk turn-off stars.  Of the
800 objects selected in the first step, 291 are kept here (63.6\% are
rejected) assuming a dust-free atmosphere, and 110 out of 357 are kept
(69.2\% are rejected) when using the model for a dusty atmosphere.

\subsection{\label{selection-3rd} Third candidate selection step: 
  Rejection based on
  observed magnitude vs.\ predicted magnitude discrepancy}

\begin{figure}
  \centering
  \includegraphics[width=8cm]{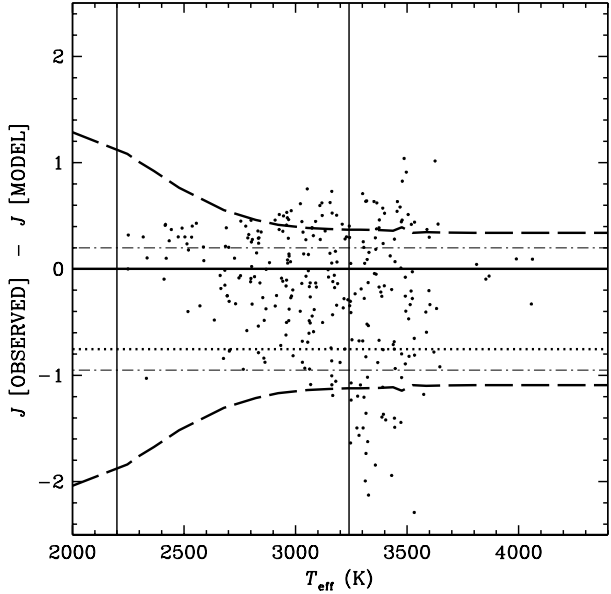}
  \caption{\label{fig:mj_vs_mj} Difference between the observed $J$
    magnitude and the model $J$ magnitude computed from the derived
    mass and $T_{\rm eff}$, as a function of $T_{\rm eff}$. The two
    vertical lines are at the positions of L0 and M5 dwarfs (left to
    right), for orientation purposes. The dotted line (at $-0.753$\,mag) represents the offset
    due to the possible presence of unresolved binaries, the
    dot--dash lines represents the error on the magnitude
    determination, and the curved, long--dashed lines represent the
    uncertainties on the age and distance of Praesepe.}
\end{figure}

As indicated in section \ref{get-mass}, our determination of $T_{\rm
  eff}$ is based on the spectral energy distribution of each object and is
independent of the assumed distance. The membership status of an object can therefore
be assessed
by comparing its observed magnitude in a band with its
magnitude predicted from its $T_{\rm eff}$ and Praesepe's isochrone (which assumes a distance).
The premise is that the predicted magnitude of a background
contaminant would be lower (brighter) than its observed magnitude and
higher (fainter) for a foreground contaminant. In order to avoid
removing unresolved binaries that are real members of the cluster, we
keep all objects with a computed magnitude of up to 0.753\,mag
brighter than the observed magnitude. We also take into account
photometric errors and uncertainties in the age and distance of
Praesepe. This selection procedure is illustrated in Figure
\ref{fig:mj_vs_mj}.  From 291 objects selected through CMDs and
colour-colour diagrams in the first two steps, 144 are kept (50.5\% are rejected) when using
the dust-free atmospheres/models, and 35 out of 110 are kept (68.2\%
are rejected) when using the dusty atmosphere/models.

After this step, we perform a visual inspection directly on the images
to reject resolved galaxies and spurious detections. This inspection
removes 21 and 8 objects from the dust-free and dusty selection
respectively.

\section{\label{results-survey} Results of the survey}

We now present the selected candidates, discuss contamination by
cluster non-members and derive the magnitude and mass functions for
Praesepe.

\subsection{\label{nbr-objects} Selected photometric candidates}

The final selection reveals 123 photometric candidates using an
isochrone based on dust-free atmospheres, and 27 objects using an
isochrone assuming dusty atmospheres\footnote{Two objects in the
  \textit{dust-free atmosphere selection} (PRAESEPE-089 and -093) were
  also identified in \textit{dusty atmosphere selection} (PRAESEPE-915
  and -917).}.  This corresponds to $\sim$40 and $\sim$9 objects per deg$^2$
respectively.  All our photometric candidates are
presented in Table \ref{tab:allphot}.  Objects are given the notation
PRAESEPE-YYY where YYY is a serial identification number (ID).
Numbers above 900 indicate candidate members assuming a dusty
atmosphere. Only the first 10 rows of the tables are shown, all other
data are available online.  We also note in Table
\ref{tab:allphot-match} which objects are candidate cluster members
also identified as such by \cite{kraus2007}, \cite{adams2002a} or
\cite{pinfield97}.

\begin{table*}
  \begin{minipage}[t]{\textwidth}
    \caption{All photometric cluster member candidates of our survey.
      Table \ref{tab:allphot} is published in its entirety in the
      electronic edition of \textit{Astronomy \& Astrophysics}.  A
      fraction is shown here for guidance regarding its form and
      content.}
    \label{tab:allphot}
    \centering
    \renewcommand{\footnoterule}{} 
    \begin{tabular}{ccccccccc}
      \hline\hline
      ID & $\alpha$ & $\delta$ & $I_{\rm c}$\footnote{The 1$\sigma$ uncertainty in the determination of magnitude, effective
        temperature and mass are the following~:
        $\Delta$mag\,=\,0.002\,mag, $\Delta$$T_{\rm eff}$\,=\,140\,K
        and $\Delta$M\,=\,0.1\,M$_\odot$ for stars
        (M\,$>$\,0.2\,M$_\odot$), $\Delta$mag\,=\,0.01\,mag,
        $\Delta$$T_{\rm eff}$\,=\,230\,K and
        $\Delta$M\,=\,0.05\,M$_\odot$ for very low-mass stars
        (0.072\,$_\odot$\,$<$\,M\,$<$\,0.2\,M$_\odot$),
        $\Delta$mag\,=\,0.04\,mag, $\Delta$$T_{\rm eff}$\,=\,420\,K
        and $\Delta$M\,=\,0.02\,M$_\odot$ for BDs
        (M\,$<$\,0.072\,M$_\odot$). The magnitude $J_{\rm model}$ is
        the predicted magnitude based on photometric determination of
        $T_{\rm eff}$ and mass.} & $J$$^a$ & $K_{\rm s}$$^a$ & M$^a$ & $T_{\rm eff}$$^a$ & $J_{\rm model}$$^a$ \\
      & ($^h$\,$^m$\,$^s$) & ($^\circ$\,$'$\,$''$) & [mag] & [mag] & [mag] & [M$_\odot$] & [K] & [mag] \\
      \hline
      001&08 40 53.61&+19 40 58.6 &19.19 &16.81 &15.61 &0.089&2665&17.00 \\
      002&08 41 08.8 &+19 43 27.5 &16.14 &14.95 &13.97 &0.219&3321&14.86 \\
      003&08 41 01.6 &+19 52 02.5 &16.67 &15.35 &14.38 &0.162&3189&15.48 \\
      004&08 41 12.17&+19 52 48.6 &18.43 &16.39 &15.46 &0.099&2805&16.72 \\
      005&08 41 08.5 &+19 54 02.0 &19.02 &16.58 &15.39 &0.088&2636&17.06 \\
      006&08 40 10.74&+19 40 49.8 &16.97 &15.47 &14.36 &0.132&3061&15.95 \\
      007&08 39 39.56&+19 47 54.3 &17.95 &16.10 &15.07 &0.104&2860&16.58 \\
      008&08 39 43.38&+19 48 45.7 &16.89 &15.56 &14.68 &0.161&3186&15.50 \\
      009&08 39 55.84&+19 53 14.3 &20.29 &17.50 &16.54 &0.081&2520&17.32 \\
      010&08 39 06.9 &+19 47 08.0 &16.51 &15.14 &14.20 &0.155&3166&15.57 \\
      \hline
    \end{tabular}
  \end{minipage}
\end{table*}

\begin{table}
  \begin{minipage}[t]{\columnwidth}
    \caption{Photometric candidates in our survey that are also
      photometric candidates in previous surveys.}
    \label{tab:allphot-match}
    \centering
    \renewcommand{\footnoterule}{} 
    \begin{tabular}{ccclc}
      \hline\hline
      ID & $\alpha$ & $\delta$ & Alternative name & Ref.\footnote{Objects [1] are from \cite{adams2002a}, 
        [2] are from \cite{kraus2007} and [3] are from \cite{pinfield97}.}\\
      & ($^h$\,$^m$\,$^s$) & ($^\circ$\,$'$\,$''$) & & \\
      \hline
      005&08 41 08.5 &+19 54 02.0 &RIZpr18&[3]\\
      010&08 39 06.9 &+19 47 08.0 &J0839069+194708&[1]\\
      & & &J08390695+1947080&[2]\\
      011&08 38 55.46&+19 50 33.3 &J0838554+195033&[1]\\
      & & &J08385547+1950334&[2]\\
      012&08 38 54.19&+19 51 44.6 &J08385420+1951446&[2]\\
      & & &J0839030+192415&[1]\\
      & & &J08390308+1924155&[2]\\
      015&08 39 12.71&+19 30 16.8 &J08391272+1930169&[2]\\
      016&08 39 54.39&+19 27 37.1 &J0839544+192737&[1]\\
      & & &J08395441+1927372&[2]\\
      017&08 39 47.82&+19 28 03.1 &RIZpr11&[3]\\
      029&08 42 50.50&+20 20 03.8 &J0842505+202004&[1]\\
      & & &J08425052+2020039&[2]\\
      034&08 42 54.58&+20 03 36.3 &RIZpr23&[3]\\
      035&08 42 51.96&+20 05 19.4 &J0842519+200519&[1]\\
      038&08 43 10.76&+20 01 29.3 &J0843107+200129&[1]\\
      040&08 41 11.05&+20 22 38.4 &J0841110+202238&[1]\\
      042&08 40 10.59&+20 20 50.4 &J0840106+202050&[1]\\
      & & &J08401060+2020505&[2]\\
      045&08 39 14.51&+20 01 19.1 &J0839145+200119&[1]\\
      046&08 39 22.43&+20 04 54.6 &J0839224+200454&[1]\\
      & & &J08392244+2004548&[2]\\
      047&08 38 55.15&+20 13 08.8 &J0838551+201308&[1]\\
      & & &J08385517+2013089&[2]\\
      054&08 40 53.96&+20 05 24.3 &J08405397+2005243&[2]\\
      062&08 36 39.46&+20 22 33.8 &J08363947+2022339&[2]\\
      064&08 36 44.99&+20 08 45.7 &J08364501+2008459&[2]\\
      066&08 37 11.41&+20 13 45.8 &J08371143+2013459&[2]\\
      068&08 38 08.0 &+20 03 50.1 &J08380800+2003505&[2]\\
      070&08 38 12.44&+20 08 02.5 &J08381244+2008026&[2]\\
      073&08 38 21.85&+20 05 35.7 &J08382186+2005356&[2]\\
      075&08 38 39.27&+19 41 40.1 &J08383929+1941401&[2]\\
      081&08 37 24.48&+19 47 11.9 &J08372449+1947120&[2]\\
      082&08 37 02.1 &+19 52 07.3 &RIZpr2 &[3]\\
      101&08 41 20.32&+18 57 42.9 &J08412034+1857430&[2]\\
      103&08 42 19.21&+19 02 14.8 &J08421923+1902148&[2]\\
      108&08 43 09.0 &+19 43 11.9 &J08430905+1943119&[2]\\
      109&08 43 01.2 &+19 49 59.8 &RIZpr24&[3]\\
      117&08 42 11.47&+19 52 50.2 &RIZpr21&[3]\\
      110&08 43 01.9 &+19 54 04.5 &J08430186+1954046&[2]\\
      112&08 42 52.26&+19 51 45.9 &J08425228+1951460&[2]\\
      116&08 42 15.48&+19 48 57.6 &J08421550+1948576&[2]\\
      122&08 43 08.4 &+19 28 06.1 &J08430839+1928061&[2]\\
      123&08 43 12.63&+19 34 28.9 &J08431265+1934290&[2]\\
      \hline
    \end{tabular}
  \end{minipage}
\end{table}

Some Praesepe members from previous studies are not detected in our
work. This is the case for the objects from \cite{pace2008} and
\cite{fossati2008}, for example.  Since those studies focused on bright
objects, these stars saturate in our science images. 
(\citealt{pace2008} and \citealt{fossati2008} were concerned with chemical
abundances of A-type and solar-type stars, respectively, while our
saturation occurs at $\sim$0.7\,M$_{\odot}$.)

Not all objects identified by other surveys as brown dwarfs or very
low mass stellar member candidates -- and detected in our survey -- 
are members based on our criteria. The two objects from the work of
\cite{gonzalez-garcia2006}, who also used photometry in order to
select candidate members, we detect above our 5$\sigma$ limit
(Prae~J084039.3+192840 and Prae~J084130.4+190449). Yet both objects are
non-members based on our selection criteria, because they have $I_{\rm
  c}-J$ colours bluer than our selection band. (Prae~J084130.4+190449
is also too blue in $I_{\rm c}-K_{\rm s}$ for our selection band at
$I_{\rm c}-K_{\rm s}$\,=\,3.0\,mag, whereas Prae~J084039.3+192840 at
$I_{\rm c}-K_{\rm s}$\,=\,4.0\,mag lies within it.)
\cite{gonzalez-garcia2006} did not report any NIR photometry for
these two objects. Although the non-membership of
Prae~J084039.3+192840 can be debated (high membership probability
based on \citealt{gonzalez-garcia2006}), Prae~J084130.4+190449 is most
likely an unresolved galaxy (low membership probability;
\citealt{gonzalez-garcia2006}).

Of the candidates from the photometric survey of
\cite{pinfield97}, seven fall within our survey and are detected, of which six are
identified as candidates by our selection criteria.  The non-selected
object is RIZpr6 in \cite{hodgkin99}. It is bluer than the isochrones
in both CMDs in Figure \ref{fig:cmd-ij} and \ref{fig:cmd-ik}. From its
positions in the CMDs and in the colour-colour diagram in Figure
\ref{fig:ccd1}, we suspect that this object is an unresolved galaxy.

11 of the the 14 objects from a survey based on proper motion and
photometry by \cite{adams2002a} are identified by our selection. The
objects not recovered fail the observed magnitude vs.\ predicted
magnitude test. On the other hand, 27 cluster candidates of
\cite{kraus2007} out of 37 detected in our survey are selected.  The
10 non-selected objects have membership probabilities from
\cite{kraus2007} based on proper motion greater than 95\%, and are
brighter than the 10$\sigma$ detection limit of the publicly available
surveys used in their work. However, these objects failed our observed
magnitude vs.\ predicted magnitude test and some are bluer than our
isochrone of Praesepe in $J-K_{\rm s}$. With $I_{\rm c}-K_{\rm s}$
colour of $\sim$\,2\,mag, we suggest that these objects are more likely to be disk
late-type stars or giant stars.

\begin{figure}
  \centering
  \includegraphics[width=8cm]{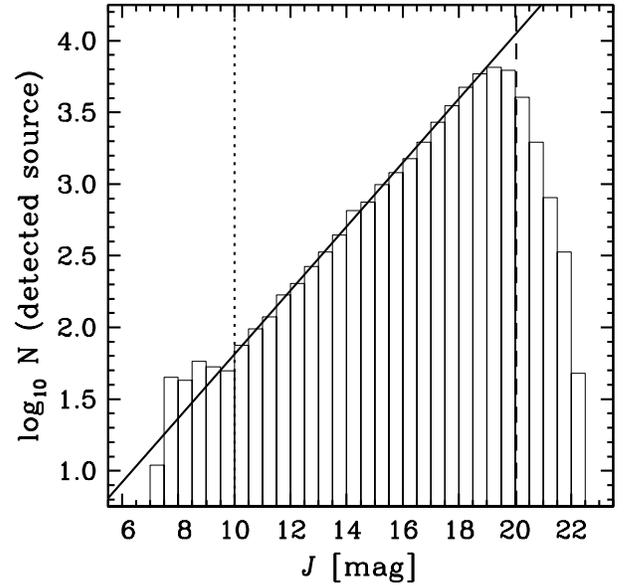}
  \caption{\label{fig:completeness} Estimation of the completeness
    limit for our survey using the $J$ band. The solid line is the
    best linear fit before the turn off, the vertical dashed line is
    the 5\,$\sigma$ detection limit and the vertical dotted line is
    the magnitude at which detector saturation occurs.}
\end{figure}

The 5$\sigma$ detection limits of our survey are $I_{\rm
  c}$=23.4\,mag, $J$=20.0\,mag and $K_{\rm s}$=18.6\,mag (which
correspond to $\sim$0.05\,M$_\odot$ using the dust-free
isochrone). However, we cannot expect to detect \textit{all} objects
down to these magnitudes. 
We estimate the survey completeness
by taking the ratio of the number of objects
detected to the predicted number of detections, the latter made by assuming a uniform
distribution of stars along the line of sight in our survey fields. 
(This comparison distribution is somewhat crude, but it gives
an approximate value without making too many assumptions.)  The
predicted number of detections is obtained from the histogram of the
number of detections as a function of magnitude (Figure
\ref{fig:completeness}) and by observing where the distribution drops off compared to a straight
line extrapolation. Based on this, the completeness of the survey down to the
5$\sigma$ detection limit is 90\% in $I_{\rm c}$, 88\% in $J$ and 87\%
in $K_{\rm s}$.  The overall detection completeness of our
survey, from saturation to 5$\sigma$ detection corresponding to
0.05\,M$_\odot$, is therefore $\sim$87\%.  In $J$ band, we reach a completeness
of 95\% at $J$=19.7\,mag, which corresponds to $\sim$0.055\,M$_\odot$.

\subsection{\label{bd-praesepe} Substellar candidates in Praesepe}

\begin{figure}
  \centering
  \includegraphics[width=8.0cm]{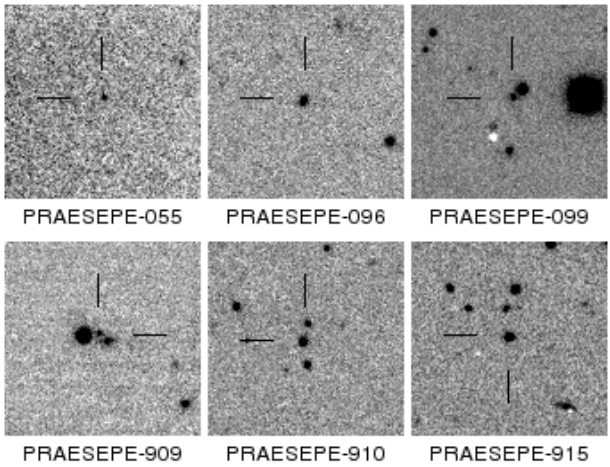}
  \caption{\label{fig:praesepe-bd} Finding charts of the six new BD
    candidates of Praesepe ($J$-band).  We observed objects very close
    to PRAESEPE-099 and -909, although they do not influence the
    photometry. The panels are 50$\times$50 arcsec with North up and
    East to the left.}
\end{figure}

Six objects in our survey are cluster candidates with theoretical
masses equal to or below the stellar/substellar boundary at
0.072\,M$_\odot$. We present the finding charts of the six objects in
Figure \ref{fig:praesepe-bd}. In Table \ref{tab:allphot-bd}, we
present their coordinates and physical parameters. These BD candidates
have predicted masses between 0.064 and 0.072\,M$_\odot$.  A
spectroscopic follow up (on a 8\,m class telescope or larger) will be
needed in order to confirm or refute their membership and their
substellar status.

\begin{table*}
  \begin{minipage}[t]{\textwidth}
    \caption{Same as Table \ref{tab:allphot}, but only the BD
      candidates are given and we include the spectral type expected.}
    \label{tab:allphot-bd}
    \centering
    \renewcommand{\footnoterule}{} 
    \begin{tabular}{cccccccccc}
      \hline\hline
      ID & $\alpha$ & $\delta$ & $I_{\rm c}$ & $J$ & $K_{\rm s}$ & M & $T_{\rm eff}$ & $J_{\rm model}$ & SpT\footnote{Spectral type expected based on $T_{\rm eff}$ and the temperature scale of \cite{kraus2007}, with L1 set to 2100\,K. The error on this estimation is one subclass.}\\
      & ($^h$\,$^m$\,$^s$) & ($^\circ$\,$'$\,$''$) & [mag] & [mag] & [mag] & [M$_\odot$] & [K] & [mag] & \\
      \hline
      055&08 41 04.5 &+20 14 58.0 &21.61 &18.29 &17.12 &0.068&2250&17.97 &L0\\
      096&08 41 13.48&+18 59 05.1 &21.06 &17.85 &16.82 &0.072&2335&17.75 &M9\\
      099&08 41 45.16&+19 18 07.7 &21.30 &17.98 &17.01 &0.068&2249&17.98 &L0\\
      909&08 39 29.94&+20 11 40.3 &20.11 &17.63 &16.67 &0.069&2259&17.95 &L0\\
      910&08 40 34.00&+20 14 56.2 &20.08 &17.60 &16.65 &0.069&2261&17.94 &L0\\
      915&08 38 51.77&+19 00 21.6 &20.28 &17.67 &16.68 &0.068&2238&18.01 &L0\\
      \hline
    \end{tabular}
  \end{minipage}
\end{table*}

\subsection{\label{contamination-rate} Contamination by non-members}

As mentioned in section \ref{selection-2nd}, the two main sources of
contamination are the background red giants, which are the dominant
source at masses of 0.09--0.2\,M$_{\odot}$ and $\sim$0.7\,M$_\odot$,
and unresolved galaxies, mostly affecting masses above
0.6\,M$_{\odot}$. Other possible contaminants are field M
dwarfs and high redshift quasars (for instance at $z$\,$\sim$\,6;
\citealt{caballero2008}). However, as such quasars have spectral
energy distributions similar to mid-T dwarfs whereas our faintest
candidates are early L dwarfs, and given that they are rare (3.3
quasars at 5.5\,$<$\,$z$\,$<$\,6.5 in a 8\,deg$^2$ survey,
\citealt{stern2007}), the MF should not be affected by quasar
contamination.

Let us estimate the contamination by M dwarfs,
First, we consider that close to the open cluster Praesepe, the space
density of M dwarfs is \textit{uniform}.  We assume that their
density ($\rho$) drops exponentially with vertical distance
from the galactic disk ($h$) such that
\begin{equation}
  \rho(h)=\rho_0e^{-\frac{h}{h_0}}, 
\end{equation} 
\noindent assuming a scale height of $h_0$\,=\,500\,pc. We use the
local space density ($\rho_0$) for M dwarfs of
57$\cdot$10$^{-3}$\,pc$^{-3}$ (from the \textit{Research Consortium on
  Nearby Stars}; \citealt{henry2006}). Given the Galactic latitude of
Praesepe of $b$=+32.5\,deg and its distance of 190\,pc, the density of
M dwarfs near Praesepe should be 47$\cdot$10$^{-3}$\,pc$^{-3}$. If we
define a volume corresponding to the area of our survey (3.1\,deg$^2$,
34\,pc$^2$) and use the distance uncertainties to the cluster
(190$^{+6.0}_{-5.8}$\,pc) as its depth, we estimate that we have
$\sim$19 M dwarf contaminants near the cluster. From a total of 150
photometric candidates, we estimate a contamination of $\sim$13\% (or
even less, as the cluster depth is presumably closer to
$\sqrt{34}=5.8$\,pc than to the 11.8\,pc error span of the mean cluster
distance). Therefore, we do not expect contamination by field M dwarfs
to play a significant role in the determination of the MF.

\subsection{\label{mf} Luminosity function and mass function}

\begin{figure}
  \centering
  \includegraphics[width=8cm]{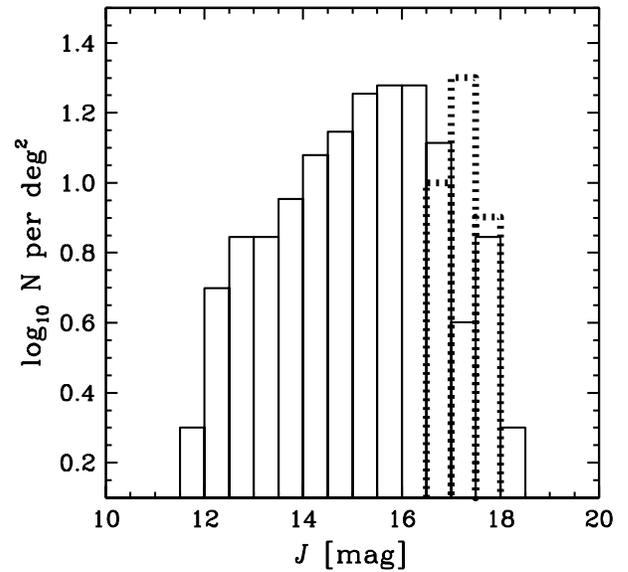}
  \caption{\label{fig:lf-prae-us} $J$ band luminosity function. The
    solid line histogram represents the luminosity function based on a
    selection using a dust-free atmosphere (NextGen model); the
    thick dotted histogram uses a dusty atmosphere (AMES-Dusty
    model). The stellar/substellar limit is at $J$$\sim$17.8\,mag for
    both models (0.072\,M$_\odot$). For reference, the ordinate value
    of 1.11 at the histogram peak (magnitude bin $J$\,=\,15.25\,mag)
    corresponds to 13 objects.}
\end{figure}

We present in Figure \ref{fig:lf-prae-us} the luminosity function of
Praesepe using the $J$-band magnitude of the cluster candidates. No
correction is made for binaries, so this is the system rather than
single-star luminosity function.

\begin{figure}
  \centering
  \includegraphics[width=8cm]{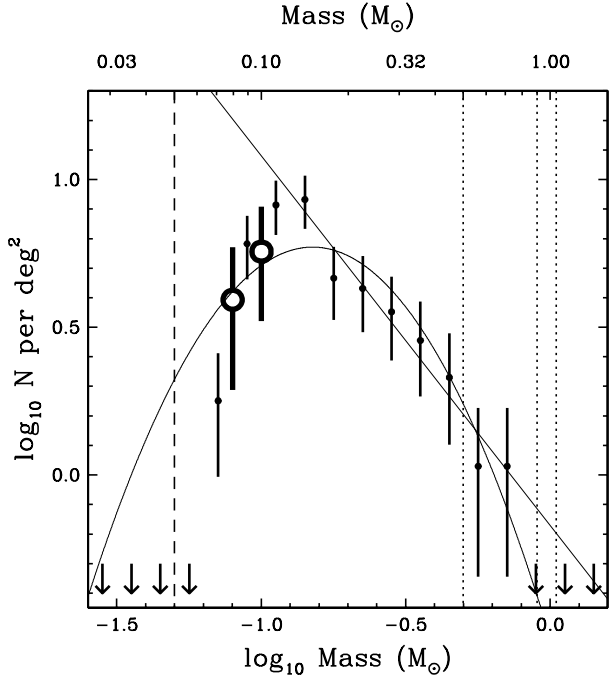}
  \caption{\label{fig:mf-prae-us} Mass function based on our survey photometry.
    Points with error bars represent the MF based on a selection and
    mass calibration assuming a dust-free atmosphere, whereas the open
    circles with error bars are the MF based on the dusty atmosphere
    model. We also overplot the log-normal and the power law MF fitted
    to our data (both solid line).  Error bars are Poissonian arising
    from the number of objects observed in each bin. The vertical thin
    dotted lines are the mass limits at which detector saturation occurs
    in the $I_{\rm c}$, $J$ and $K_{\rm s}$--bands (from
    left to right).  The
    vertical thick dashed line is the mass at the 5$\sigma$ detection
    limit (completeness of $\sim$87\%).  For reference, the ordinate
    value of 0.932 at the histogram peak (mass log$_{10}$M=$-$0.85
    [0.142\,M$_\odot$]) corresponds to 27 objects.  The two dusty data
    points have been shifted to the right by log$_{10}$M=0.05 for
    clarity.}
\end{figure}

The mass function (MF), $\xi$(log$_{10}$M), is generally defined as
the number of stars per cubic parsec in the logarithmic mass interval
log$_{10}$M to log$_{10}$M + $d$log$_{10}$M. Here, we do not compute
the volume of Praesepe so instead we define the MF as the total
number of objects in each 0.1 log$_{10}$M bin per square degree.
Since we do not make any corrections for
binaries we compute here a \textit{system} MF.  Our inferred MF is
shown in Fig.~\ref{fig:mf-prae-us}. The log-normal form for a MF is

\begin{equation}
  \xi(\textrm{log}_{10}\,\textrm{M})=k\cdot\textrm{exp}{\biggl[-\frac{(\textrm{log}_{10}\,\textrm{M}-\textrm{log}_{10}\,\textrm{M}_0)^2}{2\sigma^2}\biggr]},
\end{equation}

\noindent where $k$=0.086, $m_0$=0.22\,M$_\odot$ and $\sigma$=0.57 for
the Galactic field system MF (\citealt{chabrier2003}). Fitting this
function to both the \textit{dust-free} and \textit{dusty} MF data
we obtain $k$=5.9$\pm$3.1, $m_0$=0.15$\pm$0.05\,M$_\odot$ and
$\sigma$=0.51$\pm$0.12.  Figure \ref{fig:mf-prae-us} shows this
result. If we instead fit a power law (\cite{salpeter1955})

\begin{equation}
  \label{power-law}
  \xi(\textrm{log}_{10}\,\textrm{M})=k\cdot\textrm{M}^{-\alpha},
\end{equation}

\noindent from the highest mass bin to the turn over at
0.1\,M$_\odot$, we obtain $\alpha$=1.3$\pm$0.2 (corresponding to
$\xi(\textrm{M}) \propto M^{-2.3}$). If we exclude the two
bins between 0.1 and 0.15\,M$_\odot$ (possible contamination by red giants)
and the two highest bins (possible incompleteness), the fit gives
$\alpha$=0.8$\pm$0.1.

\section{\label{mf-variation} Analysis and discussion of the stellar
  and substellar mass function of Praesepe}

\begin{figure*}
  \centering
  \includegraphics[width=14cm]{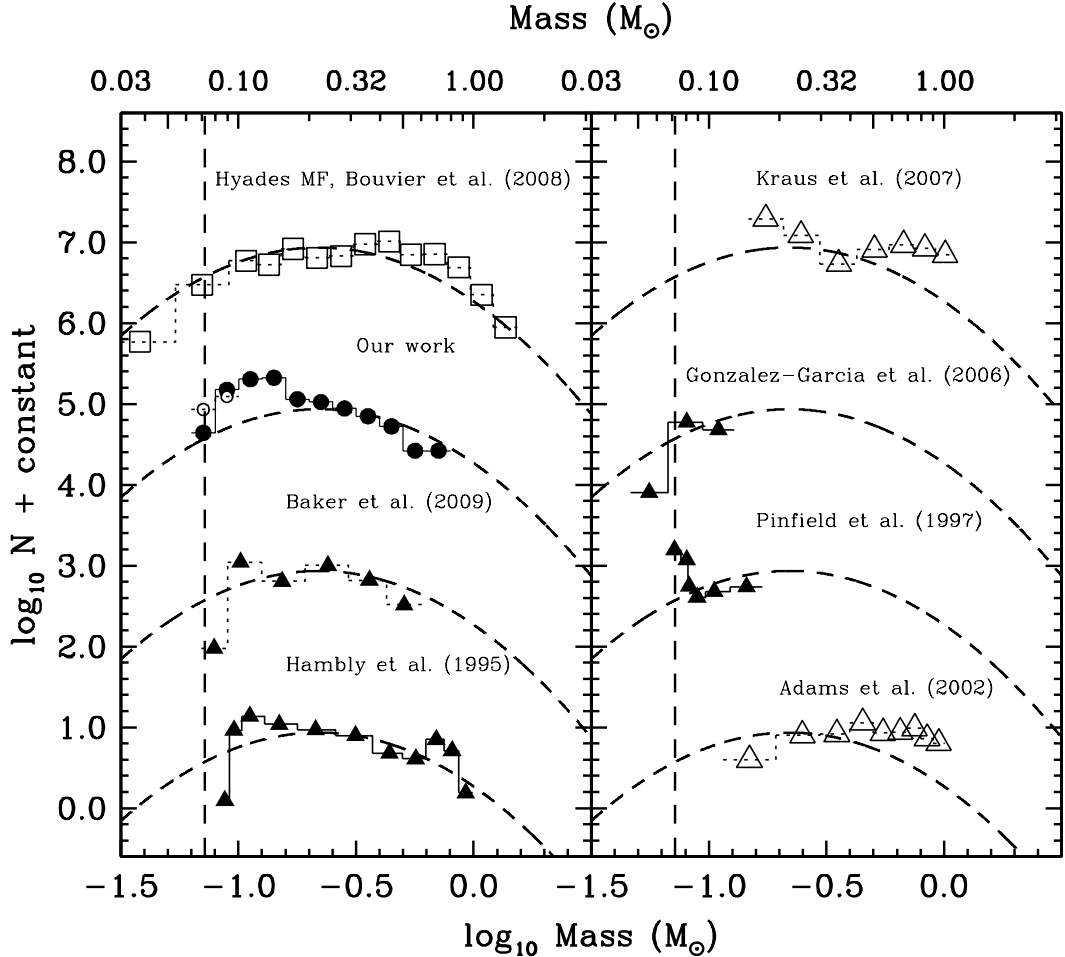}
  \caption{\label{fig:mf-all-praesepe} MF of Praesepe from our present
    work (\textit{open dots} assuming a dusty atmosphere and
    \textit{filled dots} assuming a dust-free atmosphere), from
    previous work (\textit{open triangles} for survey using proper
    motion and \textit{filled triangles} for survey using photometry
    only), as well as the MF from the Hyades from \cite{bouvier2008}
    (\textit{open squares}). We also show the galactic field star MF
    from \cite{chabrier2003} as a thin dashed line and the substellar
    limit as a thick dashed line. We have normalized all the MFs to
    the log-normal fit of \cite{chabrier2005} at $\sim$0.3\,M$_\odot$
    (log\,M=-0.5), except for those of \citealt{gonzalez-garcia2006}
    \citealt{pinfield97}) which have no data here.}
\end{figure*}

Our MF of Praesepe (Figure \ref{fig:mf-prae-us}) shows a rise in the
number of objects from 0.6\,M$_\odot$ down to 0.1\,M$_\odot$, and then
a turn-over at $\sim$0.1\,M$_\odot$.  This turn-over is not due to
incompleteness since it occurs well above the 5$\sigma$ detection
limit corresponding to 0.05\,M$_\odot$. This behaviour is confirmed
by the luminosity function in Figure \ref{fig:lf-prae-us} which shows a
rise from $J$=13 to 16\,mag (with candidates obtained using a dust-free
atmosphere) and a drop at $J$=17\,mag (seen with both types of candidates).
To help the analysis of these features in the mass function, we compare in
Figure \ref{fig:mf-all-praesepe} the mass functions of Praesepe obtained from several
studies plus the MF for the old open cluster Hyades (age of 625\,Myr).

The rise in our MF of Praesepe is also present in the MFs obtained in the
three previous studies of \cite{baker2009}, \cite{kraus2007} and
\cite{hambly1995}. On the other hand, we do not see this rise in the
MF of \cite{adams2002a}. However, their MF is based on objects with a
membership probability higher than only 1\% and within a radius of
3.8\,deg.  Due to use of such a low probability threshold for
selection, we expect that most of the objects used in the MF
determination are simply field stars (which is their own conclusion in
section 5.4; \citealt{adams2002a}), so further comparison is not
warranted. As for the MFs of \cite{gonzalez-garcia2006} and \cite{pinfield97},
since the highest mass bins are $\sim$0.11 and $\sim$0.15\,M$_\odot$
(respectively), the rise observed from 0.6\,M$_\odot$ to
0.1\,M$_\odot$ cannot be discussed.

Only four MFs, in addition to our work, reach masses below
0.1\,M$_\odot$: \cite{baker2009}, \cite{gonzalez-garcia2006},
\cite{pinfield97} and \cite{hambly1995}. While the MFs of
\cite{baker2009} and \cite{hambly1995} show a turn-over at
0.1\,M$_\odot$, the one obtained by \cite{pinfield97} does not.
On the contrary, it presents a sudden rise at the
stellar/substellar limit (with a ratio of $\sim$5 in the number of
objects at the mass bin at 0.07 to the number in the bin at
0.11\,M$_\odot$). They used $RIZ$ photometry for their survey, but not all
objects were observed in all bands, resulting in just
one colour available for membership determination in some cases
(\citealt{pinfield97}). From an analysis of MFs of other clusters and
using a multi-band photometric survey, \cite{boudreault2009} have
shown that use of a narrow spectral coverage with few filters can lead
to high contamination by field M dwarfs, and thus an apparent rise in
the MF in the low mass regime. We suggest that this is the reason for
the apparent rise at the low-mass end of the MF in \cite{pinfield97}
(who also noted that only one colour is available for many objects in
their two lowest bins). As for the MF of \cite{gonzalez-garcia2006},
as they only have three points we cannot comment on any possible
trend.

Although there are some discrepancies between the different MFs of
Praesepe from previous works and our MF, none agrees with the MF of
the Hyades ($\sim$625\,Myr) obtained by
\cite{bouvier2008}\footnote{Like the MF of Praesepe we present, the MF
  of the Hyades presented by \cite{bouvier2008} is a system MF (no
  correction for binaries).}, in which the MF is observed to turn-over
and decrease already at 0.35\,M$_\odot$. This is surprising, since
Praesepe and the Hyades share a comparable age, size and mass: they
have ages of 590$^{+150}_{-120}$\,Myr (\citealt{fossati2008}) and
625$\pm$50\,Myr (\citealt{bouvier2008}), tidal radii of
11.5$\pm$0.3\,pc (3.5$\pm$0.1\,deg, \citealt{kraus2007}) and 10.3\,pc
(12.5\,deg, \citealt{bouvier2008}), and masses of
550$\pm$40\,M$_\odot$ (\citealt{kraus2007}) and about 400\,M$_\odot$
(\citealt{bouvier2008}), respectively. Therefore, we can expect that
the potential well is the same (at least today). Only the metallicity
may be slightly different, assuming the most recent measurement for
Praesepe: [Fe/H]\,=\,+0.27$\pm$0.10 for the latest metallicity
measurement of Praesepe (\citealt{pace2008}) and
[Fe/H]\,=\,+0.14$\pm$0.05 for the Hyades (\citealt{bouvier2008}),
although a metallicity as low as +0.038$\pm$0.039
(\citealt{friel1992}) has been reported for Praesepe. It is unclear
how this metallicity difference could explain the significantly
different mass functions.

It is a priori possible that different binary mass fractions in
Praesepe and the Hyades could account for the difference in their
observed (i.e.\ system, rather than star) mass functions. The binary
fraction in Praesepe for different mass intervals was obtained by
\cite{pinfield03}: 17$^{+6}_{-4}$\% for 1.0--0.6\,M$_\odot$,
31$^{+7}_{-6}$\% for 0.6--0.35\,M$_\odot$, 44$\pm$6\% for
0.35--0.2\,M$_\odot$ and 47$^{+13}_{-11}$\% for 0.11--0.09\,M$_\odot$.
As for the Hyades, \cite{gizis1995} observed a binary fraction of
27$\pm$16\% for their sample of stars ($\lesssim$0.4\,$_\odot$), which
is consistent with another determination of the Hyades binary fraction 
of 30$\pm$6\% from \cite{patience1998} (for a primary mass of
$\sim$0.6--2.8\,M$_\odot$).  From these figures we see no significant difference in
the binary fractions of the two clusters (even if primarily
because the uncertainties are quite large), so this cannot be used
to explain the difference between in their mass functions.  Of course, if the typical mass
ratio in a binary system is different in the two clusters then this may be able
to account for some difference in the mass functions, but their
is also no evidence to support (or refute) this.

A distinction between the two clusters could be the spatial
distribution of the members. Indeed, \cite{holland2000} observed that
the Praesepe cluster might be composed of two merged clusters with
different ages, one main cluster of 630\,M$_\odot$ and a second
subcluster of 30\,M$_\odot$. It was even proposed that faint low-mass
members of the subcluster could appear as Praesepe brown dwarf candidates
(\citealt{chappelle2005}).  However, \cite{adams2002a} did not find
evidence of a subcluster in Praesepe. Based on the spatial
distribution of the main cluster and subcluster from
\cite{holland2000}, our survey only overlaps the main cluster.  In
addition, a collision between two clusters could not explain alone an
increase of the MF down to 0.1\,M$_\odot$, as such a collision would
rather remove low-mass member of the clusters.

By comparing the MF of the Hyades with the one of the Pleiades
($\sim$120\,Myr), \cite{bouvier2008} concluded that dynamical
evolution was responsible for the deficiency observed in the very-low
mass star and BD regime in the Hyades. However, this deficiency is not
seen in Praesepe. One possible implication is that Praesepe has been
less affected by dynamical evolution, i.e.\ evaporation of low mass
members which are expected to have higher speeds based on
equipartition of energy.  On the other hand, if dynamical evolution
{\em has} affected Praesepe in the same way, then it cannot have had
the same initial mass function and/or initial conditions as the
Hyades.  Dynamical interaction between one of these clusters and
another object (such as another open cluster in the past) could
explain the discrepancies between the two MFs.

\section{\label{conclusions} Conclusions}

We have presented the results of a survey to study the mass function
of the old open cluster Praesepe. The survey consisted of optical
$I_{\rm c}$-band photometry and NIR $J$ and $K_{\rm s}$-band
photometry with a total coverage of 3.1\,deg$^2$, down to the
substellar regime, with a 5$\sigma$ detection limit corresponding to
0.05\,M$_\odot$ (the detection completeness to this level is
$\sim$87\%).

Our final sample yields 123 photometric cluster member
candidates based on a selection assuming a dust-free atmosphere and 27
photometric cluster candidates based on a selection assuming a dusty
atmosphere. We estimate the contamination by field M-dwarfs to be
13\% or less. Among our cluster candidates, six objects have theoretical
masses equal to or less than the stellar/substellar boundary at
0.072\,M$_\odot$.

We observed that the MF of Praesepe is characterized by a rise in the
number of objects from 0.6\,M$_\odot$ down to 0.1\,M$_\odot$, followed by
a turn-over in the MF at $\sim$0.1\,M$_\odot$. The rise is in agreement with the
Praesepe MFs derived in several previous studies (\citealt{hambly1995};
\citealt{kraus2007}; \citealt{baker2009}) but disagrees with
\cite{adams2002a}.

We have compared the mass function of Praesepe with one derived for
the Hyades and have observed a significant difference: while the Hyades has
a maximum at 0.35\,M$_\odot$, Praesepe has a maximum at a
much lower mass, 0.1\,M$_\odot$. Assuming that they have similar ages
(as main sequence fitting suggests), we conclude that the clusters
either had different {\em initial} mass functions or that dynamical
interaction has modified the evolution of one or both. More
specifically, in the latter case, dynamical evaporation does not seem
to have influenced the Hyades and Praesepe in the same way. A
difference in the binary fraction or mass ratios could also cause a
difference in the mass functions, but determinations of these are not
yet precise enough to suggest any difference.

\begin{acknowledgements}
  SB and CBJ acknowledge support from the Deutsche
  Forschungsgemeinschaft (DFG) grant BA2163 (Emmy-Noether Programme) to
  CBJ. SB thanks the Calar Alto observatory staff for support and
  Kester Smith for observations performed in January 2007. We are
  grateful to the referee, Nigel Hambly, for his constructive comments
  and suggestions. We acknowledge Klaus Meisenheimer and
  Marie-H\'el\`ene Nicol for useful discussions about galaxy
  contamination. IRAF is distributed by the National Optical Astronomy
  Observatories, which are operated by the Association of Universities
  for Research in Astronomy, Inc., under cooperative agreement with
  the National Science Foundation. Some data analysis in this article
  has made use of the freely available R statistical package,
  http://www.r-project.org. This research has made use of the SIMBAD
  database, operated at CDS, Strasbourg, France.  This publication
  makes use of data products from the Two Micron All Sky Survey, which
  is a joint project of the University of Massachusetts and the
  Infrared Processing and Analysis Center$/$California Institute of
  Technology, funded by the National Aeronautics and Space
  Administration and the National Science Foundation.
\end{acknowledgements}


\begin{thebibliography}{}

\bibitem[Adams et al.(2002a)]{adams2002a} Adams, J.  D., Stauffer, J.
  R., Skrutskie, M. F., et al., 2002a, \aj, 124, 1570

  \bibitem[Adams et al.(2002b)]{adams2002b} Adams, T., Davies, M. B.,
    Jameson, R. F. \& Scally, A., 2002b, \mnras, 333, 547

\bibitem[Allard et al.(2001)]{allard01} Allard, F., Hauschildt, P.
  H., Alexander, D. R., et al., 2001, \apj, 556, 357

\bibitem[An et al.(2007)]{an2007} An, D., Terndrup, D. M.,
  Pinsonneault, M. H., et al., 2007, \apj, 655, 233

\bibitem[Baade et al.(1999)]{baade1999} Baade, D., Meisenheimer, K.,
  Iwert, O., et al., 1999, The Messenger 95, 15

\bibitem[Bailer-Jones \& Mundt(2001)]{calj2001} Bailer-Jones, C. A.
  L.  \& Mundt, R., 2001, A\&A, 367, 218

\bibitem[Bailer-Jones et al.(2000)]{calj2000} Bailer-Jones, C. A.,
  Bizenberger, P. \& Storz, C., 2000, Society of Photo-Optical
  Instrumentation Engineers (SPIE) Conference Series, 4008, 1305

\bibitem[Baker \& Jameson(2009)]{baker2009} Baker, D. E. A. \&
  Jameson, R. F., 2009, \textit{Brown Dwarfs in Praesepe: A search for
    low mass members using UKIDSS}, JENAM2009 meeting, poster no.
  3-P03

\bibitem[Barrado y Navascu\'es et al.(2004)]{barrado2004} Barrado y
  Navascu\'es, D., Stauffer, J. R. \& Jayawardhana, R., 2004, \apj,
  614, 386

\bibitem[Baraffe et al.(1998)]{baraffe1998} Baraffe, I., Chabrier, G.,
  Allard, F. \& Hauschildt, P. H., 1998, A\&A, 337, 403

\bibitem[Baraffe et al.(2002)]{baraffe2002} Baraffe, I., Chabrier, G.,
  Allard, F. \& Hauschildt, P. H., 2002, A\&A, 382, 563

\bibitem[Baumeister et al.(2003)]{baumeister03} Baumeister, H.,
  Bizenberger, P., Bailer-Jones, C. A. L., Kov{\'a}cs, Z., R{\"o}ser,
  H.-J. and Rohloff, R.-R., 2003, Society of Photo-Optical
  Instrumentation Engineers (SPIE) Conference Series, 4841, 343

\bibitem[B\'ejar et al.(2002)]{bejar2002} B\'ejar, V. J. S., Mart\'in,
  E. L., Zapatero Osorio, M. R., et al., 2001, \apj, 556, 830

\bibitem[Boesgaard \& Budge(1988)]{boesgaard1988} Boesgaard, A. M. \&
  Budge, K. G., 1988, \apj, 332, 410

\bibitem[Boudreault \& Bailer-Jones (2009)]{boudreault2009}
  Boudreault, S. \& Bailer-Jones, C. A. L., 2009, submitted to the ApJ
  (arXiv:0909.0842v1)

\bibitem[Bouvier et al.(2008)]{bouvier2008} Bouvier, J., Kendall, T.
  T., Meeus, G., et al., 2008, A\&A, 481, 661

\bibitem[Brice\~no et al.(2002)]{briceno2002} Brice\~no, C., Luhman,
  K. L., Hartmann, L., Stauffer, J. R. \& Kirkpatrick, J. D., 2002,
  \apj, 580, 317

\bibitem[Caballero et al.(2007)]{caballero2007} Caballero, J. A.,
  B\'ejar, V. J. S., Rebolo, R., et al., 2007, A\&A, 470, 903

\bibitem[Caballero et al.(2008)]{caballero2008} Caballero, J. A.,
  Burgasser, A. J. \& Klement, R., 2008, A\&A, 488, 181

\bibitem[Chabrier et al.(2000)]{chabrier2000} Chabrier, G., Baraffe,
  I., Allard, F. \& Hauschildt, P., 2000, \apj, 542, 464

\bibitem[Chabrier(2003)]{chabrier2003} Chabrier, G., 2003, \pasp, 115,
  763

\bibitem[Chabrier et al.(2005)]{chabrier2005} Chabrier, G., Baraffe,
  I., Allard, F. \& Hauschildt, P. H. 2005, preprint
  (astro-ph/0509798)

\bibitem[Chappelle et al.(2005)]{chappelle2005} Chappelle, R. J.,
  Pinfield, D. J., Steele, I. A., et al., 2005, \mnras, 361, 1323

\bibitem[Close et al.(2005)]{close05} Close, L. M., Lenzen, R.,
  Guirado, J. C., et al., 2005, Nature, 433, 286

\bibitem[Colina et al.(1992)]{colina1996} Colina, L., Bohlin, R. \&
  Castelli, F., 1996, Instrument Science Report CAL/SCS, 8, 1

\bibitem[D'Antona \& Mazzitelli(1985)]{dantona1985} D'Antona, F. \&
  Mazzitelli, I., 1985, \apj, 296, 502

\bibitem[de la Fuente Marcos \& de la Fuente Marcos(2000)]{marcos2000}
  de la Fuente Marcos, R. \& de la Fuente Marcos, C., 2000, Ap\&SS,
  271, 127

\bibitem[Fossati et al.(2008)]{fossati2008} Fossati, L., Bagnulo, S.,
  Landstreet, J., et al., 2008, A\&A, 483, 891

\bibitem[Friel \& Boesgaard(1992)]{friel1992} Friel, E. D. \&
  Boesgaard, A. M., 1992, \apj, 387, 170

\bibitem[Gizis \& Reid(1995)]{gizis1995} Gizis, J. \& Reid, I. N.,
  1995, \aj, 110, 1248

\bibitem[Gonz\'alez-Garc\'ia et al.(2006)]{gonzalez-garcia2006}
  Gonz\'alez-Garc\'ia, B. M., Zapatero Osorio, M. R., B\'ejar, V. J.
  S., et al., 2006, A\&A, 460, 799

\bibitem[Hambly et al.(1995)]{hambly1995} Hambly, N. C., Steele, I.
  A., Hawkins, M. R. S. \& Jameson, R. F., 1995, \mnras, 273, 505

\bibitem[Hambly et al.(1999)]{hambly1999} Hambly, N. C., Hodgkin, S.
  T., Cossburn, M. R. \& Jameson, R. F., 1999, \mnras, 303, 835

\bibitem[Harris et al.(1999)]{harris99} Harris, H. C., Vrba, F. J.,
  Dahn, C. C., et al., 1999, \aj 117, 339

\bibitem[Hauschildt et al.(1999a)]{hauschildt1999a} Hauschildt, P. H.,
  Allard, F. \& Baron, E., 1999a, \apj, 512, 377

\bibitem[Hauschildt et al.(1999b)]{hauschildt1999b} Hauschildt, P. H.,
  Allard, F., Ferguson, J., Baron, E. \& Alexander, D. R., 1999b,
  \apj, 525, 871

\bibitem[Henry et al.(2006)]{henry2006} Henry, T. J., Jao, W.-C.,
  Subasavage, J. P., et al., 2006, \aj, 132, 2360

\bibitem[Hester et al.(1996)]{hester1996} Hester, J. J., Scowen, P.
  A., Sankrit, R., et al., 1996, \aj, 111, 2349

\bibitem[Hillenbrand \& Carpenter(2000)]{hillenbrand2000}
  Hillenbrand, L. A.  \& Carpenter, J. M., 2000, \apj, 540, 236

\bibitem[Hodgkin et al.(1999)]{hodgkin99} Hodgkin, S. T., Pinfield, D.
  J., Jameson, R.  F., et al., 1999, \mnras, 310, 87

\bibitem[Holland et al.(2000)]{holland2000} Holland, K., Jameson, R.
  F., Hodgkin, S., et al., 2000, \mnras, 319, 956

\bibitem[Howell(1989)]{howell1989} Howell, S. B.  1989, \pasp, 101,
  616

\bibitem[Jameson et al.(2002)]{jameson2002} Jameson, R. F., Dobbie, P.
  D., Hodgkin, S. T. \& Pinfield, D. J., 2002, \mnras, 335, 853

\bibitem[Johnstone et al.(1998)]{johnstone98} Johnstone, D.,
  Hollenbach, D. \& Bally, J., 1998, \apj, 499, 758

\bibitem[Jordi et al.(2006)]{jordi2006} Jordi, K., Grebel, E. K. \&
  Ammon, K., 2006, A\&A, 460, 339

\bibitem[Kraus \& Hillenbrand(2007)]{kraus2007} Kraus, A. L. \&
  Hillenbrand, L. A., 2007, \aj, 134, 2340

\bibitem[Kroupa \& Bouvier(2003)]{kroupa03} Kroupa, P. \& Bouvier, J.,
  2003, \mnras, 346, 369

\bibitem[Kulkarni \& Golimowski(1995)]{kulkarni1995} Kulkarni, S. \&
  Golimowski, D., 1995, News Release Number: STScI-1995-48

\bibitem[Kumar \& Schmeja(2007)]{kumar2007} Kumar, M. S. N. \&
  Schmeja, S., 2007, A\&A, 471, 33

\bibitem[van Leeuwen(2009)]{Leeuwen2009} van Leeuwen, F. 2009, A\&A,
  497, 209

\bibitem[Lodieu et al.(2007)]{lodieu2007} Lodieu, N., Dobbie, P. D.,
  Deacon, N. R., et al., 2007, \mnras, 380, 712

\bibitem[Luhman \& Rieke(1999)]{luhman1999a} Luhman, K. L. \& Rieke,
  G.  H., 1999, \apj, 525, 440

\bibitem[Luhman(2000a)]{luhman2000} Luhman, K. L., 2000a, \apj, 544,
  1044

\bibitem[Luhman(2004)]{luhman2004} Luhman, K. L., 2004, \apj, 617,
  1216

\bibitem[Mac Low \& Klessen(2003)]{maclow04} Mac Low, M.-M.  \&
  Klessen, R. S., 2004, Rev. Mod. Phys., 76, 125

\bibitem[Magazz\'u et al.(1998)]{Ma_98} Magazz\'u, A., Rebolo, R.,
  Zapatero Osorio, M. R., et al., 1998, \apj, 497, L77

\bibitem[Meisenheimer et al.(2009)]{meisenheimer09} Meisenheimer, K.,
  Wolf, C. \& Nicol, M.-H., 2009, \textit{in prep.}

\bibitem[Moraux et al.(2003)]{moraux2003} Moraux, E., Bouvier, J.,
  Stauffer, J. R. \& Cuillandre, J.-C., 2003, A\&A, 400, 891

\bibitem[Muench et al.(2003)]{muench2003} Muench, A. A., Lada, E. A.,
  Lada, C. J., et al., 2003, \aj, 125, 2029

\bibitem[Nicol(2009)]{nicol09} Nicol, M.-H., \textit{PhD thesis},
  University of Heidelberg (2009)

\bibitem[Pace et al.(2008)]{pace2008} Pace, G., Pasquini, L. \&
  François, P. 2008, A\&A, 489, 403

\bibitem[Patience et al.(1998)]{patience1998} Patience, J., Ghez, A.
  M., Reid, I. N., et al., 1998, \aj, 115, 1972

\bibitem[Pickett et al.(2003)]{pickett03} Pickett, Brian K., Durisen,
  Richard H., Cassen, Patrick \& Mejia, Annie C., 2000, \apj, 540, L95

\bibitem[Pinfield et al.(1997)]{pinfield97} Pinfield, D. J., Hodgkin,
  S. T., Jameson, R. F., et al., 1997, \mnras, 287, 180

\bibitem[Pinfield et al.(2003)]{pinfield03} Pinfield, D. J., Dobbie,
  P.  D., Jameson, R. F., Steele, I. A., Jones, H. R. A. \&
  Katsiyannis, A.  C. 2003, \mnras, 342, 1241

\bibitem[Reid \& Hawley(1999)]{reid99} Reid, I. N. \& Hawley, S. L.,
  1999, \aj, 117, 343

\bibitem[Reipurth \& Clarke(2001)]{reipurth01} Reipurth, B. \&
  Clarke, C., 2001, \aj, 122, 432

\bibitem[Salpeter(1955)]{salpeter1955} Salpeter, E. E., 1955, \apj,
  121, 161

\bibitem[Slesnick et al.(2004)]{slesnick2004} Slesnick, C. L.,
  Hillenbrand, L. A. \& Carpenter, J. M., 2004, \apj, 610, 1045

\bibitem[Shu et al.(1987)]{shu87} Shu, F. H., Adams, F. C. \& Lizano,
  S., 1987, \araa, 25, 23

\bibitem[Stern et al.(2007)]{stern2007} Stern, D., Kirkpatrick, J. D.,
  Allen, L. E., et al., 2007, \apj, 663, 677

\bibitem[Taylor(2006)]{taylor2006} Taylor, B. J. 2006, \aj, 132, 2453

\end{thebibliography}
\end{document}